\documentclass{optica-article}

\journal{opticajournal}
\articletype{Research Article}

\usepackage{dsfont}
\usepackage{cleveref}
\usepackage{graphicx}
\usepackage{subcaption}

\renewcommand{\vec}[1]{\boldsymbol{\mathbf{#1}}}
\newcommand{\mat}[1]{\boldsymbol{\mathbf{#1}}}
\renewcommand{\Re}{\operatorname{Re}}
\renewcommand{\Im}{\operatorname{Im}}
\newcommand{\LDOS}{\text{LDOS}}
\newcommand{\T}{\mathrm{T}}

\newcommand{\citeasnoun}[1]{Ref.~\citenum{#1}}

\begin{document}

\title{Eigenvalue-accelerated LDOS optimization of high-$Q$ optical resonances}

\author{George Shaker,\authormark{1}
Be\~nat Martinez de Aguirre Jokisch,\authormark{2}
Pengning Chao,\authormark{1}
and Steven G. Johnson\authormark{1,*}}

\address{\authormark{1}Department of Mathematics, Massachusetts Institute of Technology, Cambridge, MA 02139, USA\\

\authormark{2}NNF Quantum Computing Programme, Niels Bohr Institute,
University of Copenhagen, Universitetsparken 5, 2100 Copenhagen, Denmark.}
\email{\authormark{*}stevenj@math.mit.edu}

\begin{abstract*}
We demonstrate a new method that yields orders-of-magnitude acceleration in inverse design (e.g.~topology optimization) of high-$Q$ resonant cavities to maximize the local density of states (LDOS), and which is also applicable to other resonant-response metrics. The key idea is that, once conventional LDOS optimization has identified a strong resonance, subsequent optimizations can exploit a fast shift-invert eigensolver to ensure that the LDOS remains centered at the resonance peak. We show that this eliminates ill-conditioning at  sharp resonances that otherwise dramatically slows LDOS (and similar) optimization for $Q \gg 100$. Our method is demonstrated by design of $Q > 10^6$ resonant cavities in 1d and 2d dielectric systems.
\end{abstract*}

\section{Introduction}\label{sec:intro}
Inverse design (large-scale optimization)~\cite{Molesky2018, Jensen2010} of resonant effects in optics (or other wave systems) often proceeds by maximizing the frequency-domain response to sources at a given target frequency $\omega_0$, but severe challenges are known to arise for optimizing long-lifetime cavities. A resonant response that is often optimized is the power expended by a dipole source~\cite{Liang2013, Wang2018,YaoBe20, Albrechtsen2022, Chao2022, Isiklar2022}, which corresponds to the local density of states (LDOS, reviewed in Sec.~\ref{sec:LDOS}), a key figure of merit for processes such as spontaneous emission~\cite{Molesky2018, Jensen2010,OskooiJo13-sources, Cisowski2025}, and which is often approximated by the Purcell factor $Q/V$ (the ratio of the mode quality factor $Q$ to an appropriate measure of modal volume $V$)~\cite{OskooiJo13-sources}. That is, one maximizes $\LDOS(\omega_0, \vec{x}_0, \vec{p})$ at a fixed frequency $\omega_0$ and dipole position $\vec{x}_0$ over some geometry and/or material parameters $\vec{p}$. (More generally, one can maximize some function of the field intensity excited by a source or incident wave to optimize other resonant processes~\cite{YaoVe22,Janssen2010, Kim2021, Mayer2015, Schubert2023,RoquesCarmes2022, Schuetz2023,Yao2023,Haeusler2022,Martinez2025,Mann2023, LinLi16,Christiansen2020, Pan2021}, as discussed below.)  Maximizing such a quantity not only captures many physical effects simultaneously (e.g.~lifetime, coupling, and spatial localization), but it is also amenable to efficient gradient-based optimization algorithms via adjoint methods~\cite{Liang2013,Hughes2018,Molesky2018} because the frequency-domain response is an easily differentiable function of material and geometric parameters. Although the resulting structures are resonant cavities characterized by high-$Q$ resonant modes~\cite{Christiansen2020, Wang2018, Albrechtsen2022, Liang2013, Isiklar2022}, the optimization process need never identify a particular resonant mode (eigensolution) to optimize, which eliminates the need for a good initial guess of the optimal structure and avoids potential non-differentiability of eigenvalues~\cite{Liang2013}. However, all such methods have been observed to suffer from a serious problem when the $Q$ (resonant lifetime) becomes sufficiently large ($Q \gg 100$): optimization convergence becomes slower and slower~\cite{Liang2013}, often requiring many thousands of iterations (Maxwell solves) even for $Q \sim 10^4$, making $Q \gtrsim 10^5$  costly to obtain by maximizing LDOS or similar.  The essential reason for this is that any field-intensity objective becomes like the ``edge of a knife'' at high~$Q$ (Sec.~\ref{sec:second_derivative}): performance falls off rapidly for perturbations that shift the resonant frequency away from $\omega_0$, but changes slowly for perturbations that keep the frequency fixed~\cite{Liang2013}. Optimization along such a sharp ridge is known to have poor performance due to the ill-conditioning of the second-derivative (Hessian) matrix~\cite{Nocedal2006}.

In this paper, we demonstrate \emph{orders-of-magnitude} acceleration in topology optimization~(TO)~\cite{Jensen2010} of high-$Q$ optical resonant responses by a new ``eigenfrequency-shifted'' approach (Sec.~\ref{sec:shifted_alg}) that compensates for the ill-conditioning arising from frequency sensitivity of high-$Q$ cavities~\cite{Liang2013}. We analyze our performance by explicit computation of the Hessian for a $\log\LDOS$ objective, and show both analytically (Sec.~\ref{sec:second_derivative}) and numerically (Sec.~\ref{sec:hessian}) that frequency shifts correspond to a large Hessian eigenvalue, $\sim Q^2$, that is eliminated by our frequency-shifted objective, explaining our improved optimization conditioning. Our algorithm itself does not require a Hessian, and is amenable to efficient adjoint computation of gradients for optimization (Appendix~A). We demonstrate our algorithm with TO of $\log\LDOS$ for 1d (Sec.~\ref{sec:1d_results}) and 2d (Sec.~\ref{sec:2d_results}) dielectric cavities, including the non-intuitive case of enhancing light intensity \emph{outside} the design region, obtaining $Q > 10^8$ in 1d and $Q > 10^6$ in 2d (for a tiny $1.5\lambda_0 \times 1.5\lambda_0$ cavity).  We also demonstrate further improvements by a ``successive enlargement'' heuristic, in which we gradually increase the diameter of the design region (Sec.~\ref{sec:1d_results}). In 1d, we observe a clear relationship between the diameter of the design region and the resulting $Q$ or LDOS, which increase exponentially in half-wave steps (Sec.~\ref{sec:1d_results}).

A key idea is that we can initialize our new ``shifted'' optimization with $\lesssim 10^3$ steps of the original ``unshifted'' method, which allows inverse design to discover a strong ($Q \gtrsim 100$) resonance frequency $\omega_*$ near $\omega_0$ for subsequent eigensolver-based frequency-shift compensation.  We can then optimize a frequency-shifted log-LDOS objective (Sec.~\ref{sec:shifted_alg}):
\begin{equation}
    \max_{\vec{p}} \quad \log \LDOS(\Re\left[\omega_*(\vec{p},\omega_0)\right], \vec{x}_0, \vec{p}) \, ,
\label{eq:shifted_LDOS}
\end{equation}
where $\omega_*(\vec{p},\omega_0)$ is the (complex) eigenfrequency closest to $\omega_0$, with a constraint to ensure that the eigenvalue doesn't drift too far from $\omega_0$ during optimization.  (We observe the logarithm to further accelerate convergence somewhat, similar to the use of $\LDOS^{-1}$ in previous work~\cite{Liang2013}; some optimization algorithms prefer objective functions to be of order unity~\cite{svanberg_mma_2007,wachter2006implementation}.)
Although our approach involves additional cost per $\vec{p}$ compared to the original ``unshifted'' methods, in order to perform a few steps of an iterative shift-invert eigensolve~\cite{bai2000templates}, we show that this is more than compensated by the accelerated $\vec{p}$ convergence. As further evidence of our eigenfrequency-shifted approach's performance relative to standard benchmarks, we show that we can narrow the gap between achievable LDOS and the theoretical upper bound~\cite{Chao2022, Chao2025} in lossy materials.

Our examples in this paper maximize LDOS, but a similar approach could be applied (Sec.~\ref{sec:conclusion}) to maximizing other frequency-domain resonant responses. For example, the LDOS has been replaced by other metrics for designing other incoherent emission processes~\cite{YaoVe22} such as light-emitting diodes~\cite{Janssen2010, Kim2021, Mayer2015, Schubert2023}, scintillation~\cite{RoquesCarmes2022, Schuetz2023}, or Raman scattering~\cite{Yao2023}.  Additional linear-response objectives have been used to optimize Smith--Purcell radiation~\cite{Haeusler2022} or lasing efficiency~\cite{Martinez2025}. One can even cascade multiple linear responses to maximize nonlinear effects such as second-harmonic generation~\cite{Mann2023, LinLi16} or Raman effects~\cite{Christiansen2020, Pan2021}. In all such cases, $\Re\left[\omega_*(\vec{p},\omega_0)\right]$ can again be employed to stay on the resonance peak and accelerate convergence after an unshifted initialization identifies a target resonance.  While we \emph{exploit} this eigenvalue, we believe that one should generally strive for a holistic physical objective like LDOS that captures as much as possible of the physics of a given problem, including both lifetime ($Q$) and localization effects (as opposed to specialized cases where one targets $Q$ without optimizing localization~\cite{Frei2008,Ahn2022,Li2023} or targets a localization metric without consideration for~$Q$~\cite{Lu2010,ChristiansenMork2024}).

\section{Local density of states (LDOS)}
\label{sec:LDOS}
In this section, we review the formulation of the time-harmonic Maxwell equations, resonant modes, the quality factor $Q$, and the local density of states (LDOS).  For illustration purposes, we consider the 2d Maxwell equations in the $xy$ plane for an out-of-plane ($E_z$) polarization, although the general principles of the LDOS also apply to the full 3d Maxwell equations (as well as to other wave equations).

For a target excitation frequency $\omega_0 > 0,$ a permittivity distribution $\varepsilon(\vec{p},\vec{x})$ for $\vec{x} = (x,y) \in \mathbb{R}^2$, and a source $J_z(\vec{x}) = \delta(\vec{x} - \vec{x}_0),$ the governing physics on the electric field $E_z(\vec{x})$ are given by the inhomogeneous Helmholtz equation
    \begin{equation}
    -\left(\nabla^2 + \omega_0^2\varepsilon\right)E_z = i\omega_0 J_z,
    \label{eq:Helmholtz_continuous}
    \end{equation}
with Sommerfeld radiation boundary conditions, for dimensionless/natural units $\varepsilon_0 = \mu_0 = c = 1$.  The corresponding vacuum ($\varepsilon = 1$) wavelength is $\lambda_0 = 2\pi /\omega_0$; below, we express distance in units of $\lambda_0$ (equivalently, $\lambda_0 = 1$ and $\omega_0 = 2\pi$). Numerically, we discretize the Helmholtz equation using finite differences on a uniform grid---a finite-difference frequency-domain (FDFD) method~\cite{Christ1987,rumpf2022electromagnetic,Shin2012}---to obtain
    \begin{equation}
    \underbrace{-\left(\mat{L} + \omega_0^2\mat{D}\right)}_{\mat{A}}\vec{e} = \vec{b},
    \label{eq:Helmholtz_discrete}
    \end{equation}
where $\mat{L}$ is a matrix approximating the second derivative, $\mat{D} = \mathrm{diag}(\vec{\varepsilon}),$ and $\vec{b}$ is a one-hot vector corresponding to a point source at $\vec{x}_0$. To implement outgoing boundary conditions, we surround the design region with perfectly matched absorbing layers (PML)~\cite{taflove_em_book}, which are accounted for in the matrix $\mat{L}$ as described in Appendix~B. Henceforth, we will denote the matrix in Eq.~\eqref{eq:Helmholtz_discrete} by $\mat{A}(\omega_0, \vec{x}_0, \vec{p}) = -\left(\mat{L} + \omega_0^2\mat{D}\right),$ omitting the arguments where convenient.  These matrices are sparse, so we solved the system $\mat{A}\vec{e} = \vec{b}$ using a sparse-direct algorithm~\cite{davis2006direct,UMFPACK}.

Electromagnetic reciprocity implies that the Helmholtz operator $\nabla^2 + \omega^2 \varepsilon$ is symmetric for scalar $\varepsilon$, even in the presence of loss or absorbing boundaries~\cite{Chew2008,Guo2022}.  It is convenient below to choose the discretization to obtain the same property $\mat{A} = \mat{A}^{\T}$ for the matrix $A$, so that we do not need to distinguish left and right eigenvectors of $\mat{A}$.  Although the typical stretched-coordinate PML formulation breaks this symmetry, the symmetry of $\mat{A}$ can be restored by a simple diagonal scaling as described in Appendix~B.

From our discretization, one may relate the quality factor to the governing physics of the Helmholtz equation with the following generalized eigenvalue problem:
    \begin{equation}
    -\mat{L}\vec{e}^{(k)} = (\mat{A} + \omega_0^2\mat{D})\vec{e}^{(k)} = \omega_k^2\mat{D}\vec{e}^{(k)}.
    \label{eq:eigenproblem}
    \end{equation}
The quality factor $Q$ of the mode $\vec{e}^{(k)}$ with eigenfrequency $\omega_k$ is defined as~\cite{JoannopoulosJo08-book,OskooiJo13-sources}
    \begin{equation}
    Q = -\frac{\Re\left[\omega_k\right]}{2\Im\left[\omega_k\right]}.
    \label{eq:quality_factor}
    \end{equation}
The LDOS at $\vec{x}_0$ is the power expended by a time-harmonic point dipole current source, equal to $-\frac{1}{2}\Re\left[\int \overline{E} \cdot J \, d\Omega\right]$ in Maxwell's equations~\cite{OskooiJo13-sources}, which in the discretized equations simplifies to:
    \begin{equation}
    \LDOS(\omega_0, \vec{x}_0, \vec{p}) \sim -\Im\left[\vec{e}^{\dag} \vec{b}\right],
    \label{eq:LDOS}
    \end{equation}
where we have dropped a factor of $1/2\omega_0$ for simplicity---typically, our objective is the dimensionless LDOS \emph{enhancement} relative to some reference system (e.g.~vacuum), in which case such scale factors cancel. Henceforth, we denote the right-hand side of this expression as the ``$\LDOS$''. It is nonnegative by conservation of energy, allowing us to equivalently work with $\log \LDOS$ or $\LDOS^{-1}$~\cite{Liang2013}.  The LDOS is computationally convenient for optimization because it can be computed from the solution of a single system  $\mat{A}\vec{e} = \vec{b}$ and is a smooth function of any smooth parameterization $\vec{p}$ of the discretized $\vec{\varepsilon}$, easily differentiable by adjoint methods~\cite{Liang2013,Hughes2018,Molesky2018,matrix_calculus} as reviewed in  Appendix~A.

In 3d, or in 2d with the $H_z$ polarization, the LDOS can diverge at sharp corners~\cite{Miller2016,Choi2017}, so to obtain a finite optimum one would need to regularize the problem~\cite{ChenCh24} by imposing a minimum lengthscale~\cite{Albrechtsen2022, Chao2022, Isiklar2022, ChenCh24}, specifying a nonzero separation between the emitter and the material~\cite{Miller2016,Chao2022}, and/or incorporating nonlocal effects~\cite{Cirac2012,Toscano2012}.  However, these concerns are independent of the ill-conditioning arising from high $Q$, and such singularities do not occur in 2d for the $E_z$ polarization, so we need not consider them here.

\subsection{$Q$ dependence of LDOS second derivatives}
\label{sec:second_derivative}
Consider a high-$Q$ resonance at a complex frequency $\omega_{*}(p)$ that depends on some parameter(s) $p$ of the geometry.  That is, $Q = -\Re\left[\omega_{*}\right]/2\Im\left[\omega_{*}\right] \gg 1$.  At frequencies $\omega$ near $\omega_{*}$, for sources that couple strongly to the resonant mode, the LDOS is dominated by the resonant response and takes the form of a sharp peak.  In particular, it is well known that the LDOS near a single strong resonance is approximately a Lorentzian peak~\cite{OskooiJo13-sources}:
    \begin{equation}
        \LDOS(\omega, p) \approx \frac{A(p) \gamma(p)}{\delta(p,\omega)^2 + \gamma(p)^2},
    \label{eq:lorentzian}
    \end{equation}
for linewidth $\gamma = -\Im\left[\omega_{*}(p)\right] = \Re\left[\omega_{*}(p)\right]/2Q$ and detuning $\delta = \omega - \Re\left[\omega_{*}(p)\right]$, with some coupling strength $A(p)$ depending on the geometry (but not depending directly on $Q$). 

For optimization over $p$, a key quantity is the second derivative\cite{Nocedal2006}, which is very large near such a sharp resonant peak, since any change in $p$ will tend to shift the resonance frequency $\Re\omega_{*}$ away from the target frequency $\omega = \omega_0$.  This can be quantified using the Lorentzian model.  One finds that the second derivative at $\delta = 0$ (the peak) with respect to a scalar $p$ is:
\begin{equation}
\left. \frac{\partial^2 \LDOS}{\partial p^2}  \right|_{\omega = \Re\omega_*} \approx \frac{A''}{\gamma} - \frac{A\gamma'' + 2A'\gamma'}{\gamma^2} + \frac{2A\left(\gamma'^2 - \delta'^2\right)}{\gamma^3}\, ,
\label{eq:d^2lorentzian}
\end{equation}
where primes denote derivatives with respect to $p$, and the dominant $-2A\delta'^2 \gamma^{-3} = O(Q^3)$ term is  precisely $\frac{\partial^2 \LDOS}{\partial \omega^2} \left(\frac{\partial \Re\omega_*}{\partial p} \right)^2$.   Applying the same analysis to $\log \LDOS$ yields a dominant term that scales as $O(Q^2)$, since one factor of $\gamma$ from the numerator separates out.  More important than the absolute magnitude of the second derivative is the \emph{ratio} of the large second derivative for frequency shifts ($\delta'$) to the small second derivative for improvements in the amplitude ($A''$), and this ratio also scales as $O(\gamma^{-2}) = O(Q^2)$, regardless of whether one looks at $\LDOS$ or $\log \LDOS$ or $\LDOS^{-1}$.  (It is such ratios that determine the conditioning and influence the convergence rate of the optimization problem~\cite{Nocedal2006}.)

Hence, we expect that the dominant term in our $\log\LDOS$ Hessian should scale as $O(Q^2)$ and arise from the dependence of the resonant frequency on the parameters. We confirm this scaling numerically in Sec.~\ref{sec:hessian} for a vector $\vec{p}$ of parameters: the dominant eigenvalue of the Hessian (second-derivative) matrix indeed scales with $Q^2$ and the corresponding eigenvector is nearly parallel to $\nabla_{\vec{p}} \Re\omega_*$ (the direction of maximal frequency shift), exactly as predicted by this simple analysis.  Thus, it is crucial to correct for resonant-frequency shifts in order to obtain a well-conditioned optimization problem.

\section{Resonance/LDOS optimization}
\label{sec:optimization}
\subsection{(Old) Unshifted algorithm}
\label{sec:unshifted_alg}   
As proposed by \citeasnoun{Liang2013} and adopted by several subsequent works~\cite{Wang2018,YaoBe20, Albrechtsen2022, Chao2022, Isiklar2022}, one can design a resonant cavity by directly maximizing the LDOS (or some monotonic function thereof, perhaps with bandwidth regularization via added loss~\cite{Liang2013}), i.e.:
\begin{equation}
\begin{aligned}
    \max_{\vec{p}} \quad & \log \LDOS(\omega_0, \vec{x}_0, \vec{p}),
\end{aligned}
\label{eq:unshifted_opt}
\end{equation}
where $\vec{p}$ is some parametrization of the geometry/materials; for example, in topology optimization (reviewed in Sec.~\ref{sec:topopt})  essentially ``every pixel'' $\varepsilon$ of the design is a degree of freedom.  As we commented for Eq.~\eqref{eq:shifted_LDOS}, we take the logarithm of the $\LDOS$ (which is always positive) because we observe this to yield slightly faster convergence, similar to the $\LDOS^{-1}$ suggested in \citeasnoun{Liang2013}, perhaps in part because many optimization algorithms favor objective functions scaled to be of order unity~\cite{svanberg_mma_2007,wachter2006implementation} (due to dimensional hyperparameter choices in the algorithms).

A key advantage of directly maximizing the LDOS is that it is a smooth function of the geometry, and does not require one to select a particular resonant mode to optimize.  In practice, one finds that maximizing LDOS (or similar figures of merit) quickly creates a resonant mode with a moderately high $Q \gg 10^2$, but then converges more slowly as $Q$ (and hence LDOS) gradually increases~\cite{Liang2013,ChenCh24}.   Examples of this behavior are given in Sec.~\ref{sec:1d_results} and Sec.~\ref{sec:2d_results}, below. In some cases, to help the LDOS optimization locate a resonant mode, it can be effective to broaden the resonances by adding artificial absorption $\alpha$ to the problem, e.g.~multiplying $\varepsilon$ everywhere by $1 + i\alpha$, and then decreasing $\alpha \to 0^+$ as the optimization progresses, or alternatively employing $\alpha > 0$ to impose a bandwidth regularization~\cite{Liang2013}.  The key fact that LDOS maximization can quickly identify/create a good-quality resonant mode near the target frequency, even starting from a poor initial guess (such as vacuum or a random structure) is what we exploit for our new ``shifted'' objective in the next section.

\subsection{(New) Shifted algorithm}
\label{sec:shifted_alg}
\begin{figure}[tbp]
    \centering
    \includegraphics[width=\linewidth]{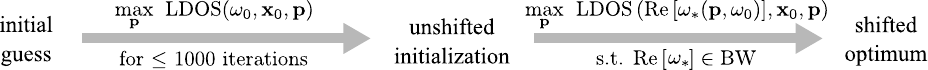}
    \vspace{5pt}
    \caption{Schematic of new eigenvalue-``shifted'' algorithm: from an initial structure (e.g.~vacuum) we run $\lesssim 1000$ iterations of the original ``unshifted'' optimization of LDOS~\cite{Liang2013} to acquire an unshifted initialization which serves as the starting guess for the shifted optimization of Eqs.~(\ref{eq:shifted_LDOS},\ref{eq:shifted_opt}): maximizing the LDOS at the real part of nearest eigenfrequency $\omega_*$, constrained within some bandwidth BW, to ensure that the objective remains on the resonance peak.}
    \label{fig:1d_roadmap}
\end{figure}

After a few steps of unshifted optimization, one typically finds that the source excites a dominant resonance with $Q \gtrsim 10^2.$ This provides us with an unambiguous resonant eigenfrequency for subsequent optimization: the eigenfrequency closest to the target $\omega_0$. Let $\omega_*(\vec{p}, \omega_0)$ denote the eigenfrequency closest to $\omega_0$ and let $\vec{e}_*$ be the associated eigenvector. Since our unshifted initialization starts with a resonance quite close to $\omega_0$ already, one can quickly compute $\omega_*$ in a few iterations of efficient methods like shift-invert Arnoldi ~\cite{bai2000templates}.  As shown in Appendix~A, this requires only one additional expensive sparse-matrix factorization compared to the unshifted LDOS, including the gradient computation, approximately doubling the cost of each optimization step. Naively, one could then directly minimize something like $|\omega_*(\vec{p}, \omega_0) - \omega_0|^2$ to obtain a high-$Q$ resonance near $\omega_0$, but that ignores the spatial-localization metrics of objectives like the LDOS. A hybrid objective~\eqref{eq:shifted_LDOS}, computing the LDOS (or similar) linear-response objective at the eigenfrequency $\Re \omega_*$, captures all of the relevant physics while improving conditioning (keeping the optimizer centered on the peak frequency) and forestalling the need to differentiate an eigenvector-dependent objective (Sec.~\ref{sec:conclusion}). As we shall show numerically in Sec.~\ref{sec:hessian}, changing the objective function to be consistently centered on $\Re\omega_*$ eliminates the $\mathcal{O}(Q^2)$ eigenvalue of the Hessian. Hence, the optimizer is able to take larger steps and converges more rapidly. Our resulting ``shifted'' optimization problem is
    \begin{equation}
    \begin{aligned}
        \max_{\vec{p}} \quad & \log \LDOS\left(\Re\left[\omega_*(\vec{p},\omega_0)\right], \vec{x}_0, \vec{p}\right)\\
    \text{s.t.} \quad & \Re\left[\omega_*(\vec{p},\omega_0)\right] \in \mathrm{BW}(\omega_0) \, .
    \end{aligned}
    \label{eq:shifted_opt}
    \end{equation}
Here, the bandwidth (BW) constraint is meant to prevent the optimizer from allowing $\omega_*$ to drift far from $\omega_0$ and to eliminate any instances of ``jumping'' to another mode. It merely serves a stabilizing purpose for the first few thousand iterations, and we found that it is not usually active throughout the optimization. Through  experimentation, we found an appropriate bandwidth constraint for the test problems in this paper was often to simply bound $\Re \omega_*$ below by $\omega_0$ or a nearby frequency, i.e. $\mathrm{BW}\left(\omega_0\right) = [\omega_0, \infty)$, although this may vary from problem to problem. We discuss the choice of constraint further in Sec.~\ref{sec:conclusion}. Since $\Re \omega_*$ and $\nabla_{\vec{p}} \Re \omega_*$ are already computed as part of the $\log\LDOS$ objective and its gradient (Appendix~A), adding one or more constraints on $\Re \omega_*$ is computationally ``free'' in that it requires no additional Maxwell solves. Fig.~\ref{fig:1d_roadmap} depicts a schematic summary of our procedure.

\section{Topology optimization (TO): Review}
\label{sec:topopt}
Having discretized the Helmholtz Eq.~\eqref{eq:Helmholtz_discrete}, we will employ density-based topology optimization (TO)~\cite{Jensen2010}, in which the parameters $\vec{p}$ describe an artificial ``density'' $\vec{\rho} \in [0,1]^N$ (one per grid point, in our case).
This density can then be linearly mapped to the corresponding permittivity $\varepsilon_k$ at each grid point $k$ in the design region ($\vec{x}_k \in \Omega$):
\begin{equation}
    \varepsilon_k = \varepsilon_{\min} + \left(\varepsilon_{\max} - \varepsilon_{\min}\right)\rho_k \, ,
    \label{eq:material_interpolation}
\end{equation}
where $\varepsilon_{\min}=1$ and $\varepsilon_{\max}=12$ are the minimum and maximum permittivities, respectively.  A key question is to precisely define how the parameters $\vec{p}$ relate to the densities $\vec{\rho}$.   Typically, one takes $\vec{p} \in [0,1]^N$ to be density degrees of freedom, but does \emph{not} directly set  $\vec{\rho} = \vec{p}$.  Instead, one low-pass filters the $\vec{p}$ in order to regularize the problem by imposing a minimum lengthscale, and then passes the resulting smoothed densities through some form of approximate Heaviside projection in order to approximately binarize the final density $\vec{\rho}$ to be nearly $0$ or $1$ almost everywhere~\cite{Jensen2010}.

In our 1d examples below, we omitted these filter/project steps and simply set $\vec{\rho} = \vec{p}$ (which is equivalent to directly optimizing over $\vec{\varepsilon} \in [\varepsilon_{\min}, \varepsilon_{\max}]$ at each grid point in the design region), similar to \citeasnoun{Liang2013}.  It turned out that filtering was not necessary to regularize or project the 1d problem, since the designs in Sec.~\ref{sec:1d_results} invariably seemed to converge to something resembling a binary ``quarter-wave stack''~\cite{JoannopoulosJo08-book} where the layer thicknesses were on the order of a quarter wavelength in $\varepsilon_{\min}$ or $\varepsilon_{\max}$.  (Similar binary structures were observed in 2d, especially for the out-of-plane polarization, by \citeasnoun{Liang2013}, and there are theoretical arguments that unconstrained cavity optimization will generally lead to binary designs~\cite{Osting2013}.)

For our 2d results (Sec.~\ref{sec:2d_results}), however, we found that unfiltered TO led to many pixel-scale features, and so we employed a filter-project algorithm using a conic filter~\cite{Jensen2010} of radius~$0.1\lambda_0$, along with a recently developed subpixel-smoothed projection (SSP)~\cite{Hammond2025} algorithm.  The projection (binarization) strength is controlled by a hyperparameter $\beta$ that is essentially the steepness of a smoothed-Heaviside step function, which is gradually increased to binarize the structure during optimization while allowing the topology to smoothly change at early stages~\cite{Jensen2010}.  SSP allows us to set $\beta$ to $\infty$ in the final optimization stage in order to guarantee a structure that is binarized except in a 1-pixel layer at interfaces, while remaining differentiable (effectively a level-set method)~\cite{Hammond2025}.   In our 2d experiments below, we increased $\beta$ through $8, 16, 40, \infty$ during the unshifted ``initialization'' stage to find a moderate-$Q$ resonance, but all subsequent 2d optimization was performed at $\beta = \infty$ for fully binarized structures.

Given an implementation of an objective function such as $f(\vec{p}) = \LDOS(\vec{p})$, here using the FDFD implementation from Sec.~\ref{sec:LDOS}, one computes the gradient $\nabla_{\vec{p}} f$ with respect to $\vec{p}$, at which point there is a wide selection of gradient-based optimization algorithms. The gradient is computed efficiently by backpropagating the derivatives through both the filtering and projection (via reverse-mode automatic differentiation~\cite{JMLR:v18:17-468,griewank2008evaluating,matrix_calculus,Zygote.jl-2018}) and through the PDE solver by manual implementation of an adjoint method~\cite{Hughes2018,Molesky2018,matrix_calculus} (described in Appendix~A).   We then employed an optimization algorithm by Svanberg~\textit{et~al.}~\cite{svanberg_class_2002,svanberg_mma_2007} as implemented by the \texttt{CCSAQ} method of the free-software NLopt package~\cite{NLopt}.

\begin{figure}[!tpb]
    \centering
    
    % Top row: (a) 1d setup
    \begin{subfigure}[t]{\textwidth}
        \centering
        \caption{}
        \includegraphics[width=\linewidth]{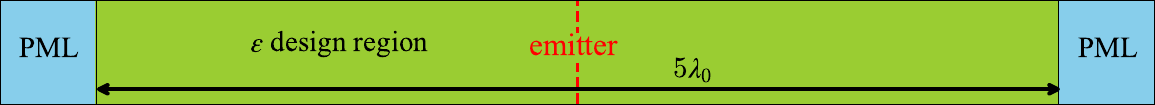}
        \label{fig:1d_setup}
    \end{subfigure}
    
    % Second row: (b) 1d cavities & modes - base algorithm
    \begin{subfigure}[t]{\textwidth}
        \centering
        \caption{}
        \includegraphics[width=\linewidth]{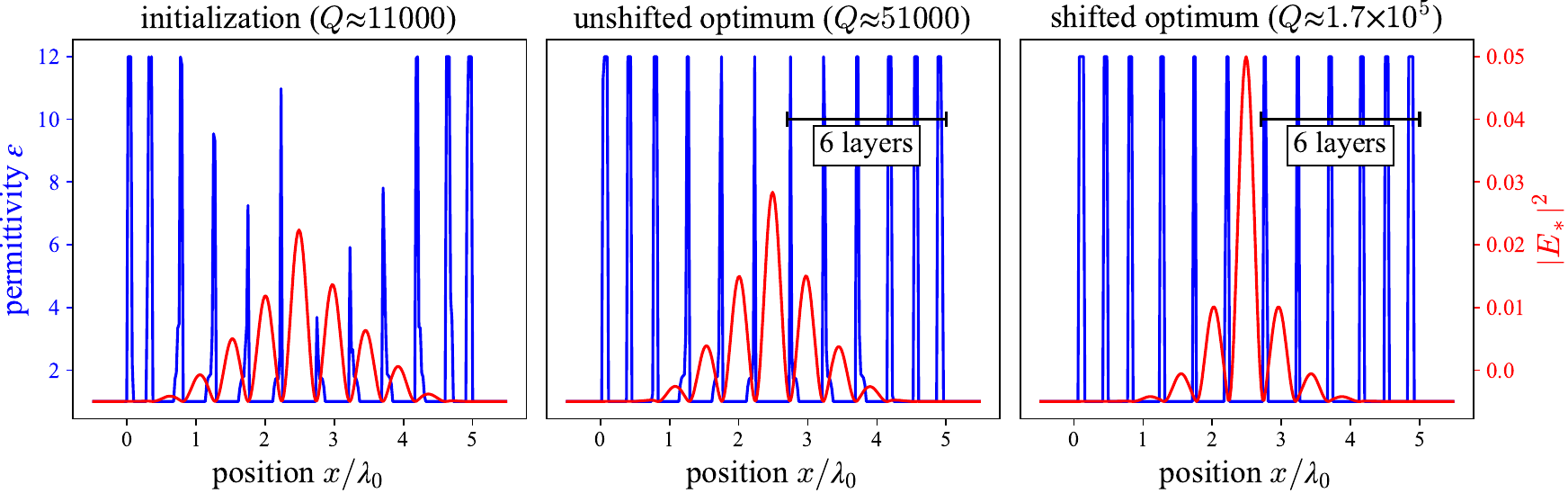}
        \label{fig:1d_cavities}
    \end{subfigure}
    
    % Third row: (c) 1d plots - base algorithm
    \begin{subfigure}[t]{\textwidth}
        \centering
        \caption{}
        \includegraphics[width=\linewidth]{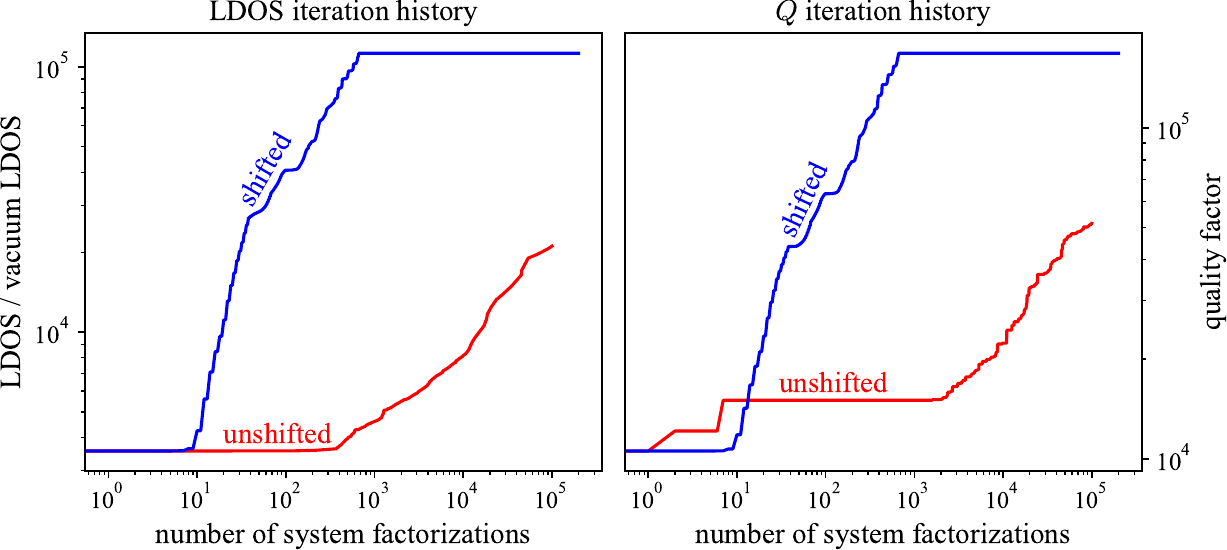}
        \label{fig:1d_plots}
    \end{subfigure}
    
    \caption{(a) Schematic 1d optimization problem: maximizing LDOS for an emitter at the center for a design region that spans the whole computational cell except for terminating PML absorbing layers. (b) Unshifted initialization after 200 iterations (left) and shifted (right) and unshifted (middle) optima after $10^5$ iterations: the resonant-mode profile (red, right axis) is superimposed over the permittivity~$\varepsilon$ (blue, left axis). (c) LDOS enhancement (left) and resonance $Q$ (right) as a function of the number of sparse-matrix factorizations (dominating the cost of the Maxwell solves) for the shifted (blue) and unshifted (red) algorithms.}
\end{figure}

\section{One-dimensional results}
\label{sec:1d_results}

To begin with, we consider a 1d problem in which the fields and materials only depend on $x$, with current $J_z = \delta(x)$.  Such geometries allow rapid computational exploration, and are also easier to understand because the maximal localization in an infinite structure is expected to arise from the bandgap of a quarter-wave stack~\cite{JoannopoulosJo08-book,Osting2013}, consisting of alternating layers of air (with thickness $\lambda_0/4$) and dielectric (with thickness $\lambda_0/4\sqrt{\varepsilon}$, here with $\varepsilon = 12$), at a given vacuum wavelength $\lambda_0$, within which a cavity is created by a ``defect'' (a perturbed layer thickness).  So, one might expect similar alternating-layer structures to arise from inverse design, with $\LDOS$ and $Q$ increasing exponentially with the diameter of the design region, albeit with some deviation from an exact quarter-wave stack due to the finite size.  In 1d, $Q > 10^8$ should be easily attainable with tractable computational domains.  However, due to ill-conditioning, the convergence rate of inverse design will become very slow for high-$Q$ structures, and should be improved by our new eigenvalue-shifted algorithm.

\subsection{Inverse design with a centered emitter}
We consider a design region $\Omega$ of width $5\lambda_0$ (terminated on both sides by PML of thickness $0.5\lambda_0$) with the source at the center of $\Omega$, as depicted in Fig.~\ref{fig:1d_setup}. Our 1d FDFD grid has resolution $50\,\text{pixels}/\lambda_0$ .

Starting from an initial $\varepsilon = 1$, we acquired an unshifted initialization, shown in Fig.~\ref{fig:1d_cavities}~(left), by running $200$ iterations of the unshifted algorithm. From this unshifted initialization, we then ran 100,000 iterations of the unshifted and shifted algorithms and found that the shifted algorithm quickly converges to a fully binary alternating-layer structure with $Q \approx  1.7 \times 10^5$, in Fig.~\ref{fig:1d_cavities}~(right), while the unshifted algorithm stalls around $Q \approx 5.1 \times 10^4$ (middle).  The convergence history of the optimization algorithm, for both $\LDOS$ and $Q$, is shown in Fig.~\ref{fig:1d_plots}, illustrating the dramatically better convergence of the shifted algorithm: after $100,000$ iterations, the unshifted algorithm has improved $\LDOS$ by half as much as the shifted algorithm, which had already converged many tens of thousands of iterations ago, and the unshifted algorithm still has not converged.  As described in Appendix~A, the shifted algorithm requires twice as many sparse-matrix factorizations as the unshifted algorithm, approximately doubling the cost of each iteration, but we account for this by plotting the convergence as a function of the number of Maxwell-matrix factorizations. Both the unshifted and shifted optimizations are converging to cavities with 6~high-index layers on either side, identical to the initialization structure---we will see below that this is, in fact, only a local optimum, and a much better optimum can be found by a more careful initialization.

\subsection{Successive enlargement}
\begin{figure}[tbp]
    \centering
    % Top row: (a) successive refinement schematic and (b) LDOS vs L
    \begin{subfigure}[t]{0.45\textwidth}
        \centering
        \caption{}
        \includegraphics[width=\columnwidth]{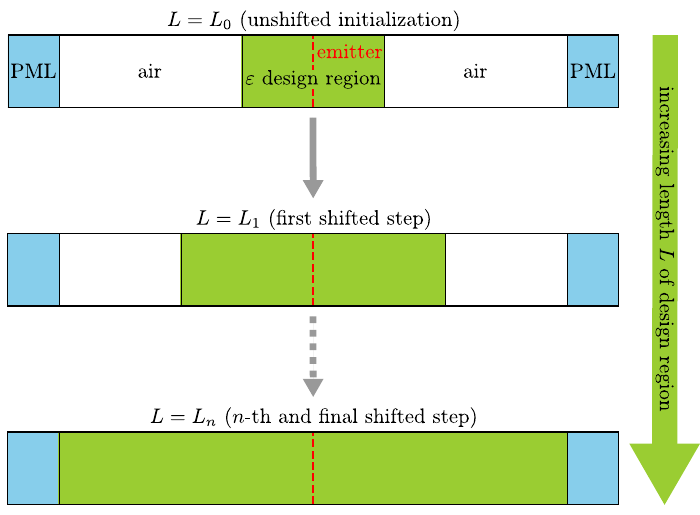}
        \label{fig:succ_ref_schematic}
    \end{subfigure}%
    \hfill
    \begin{subfigure}[t]{0.55\textwidth}
        \centering
        \caption{}
        \includegraphics[width=0.85\columnwidth]{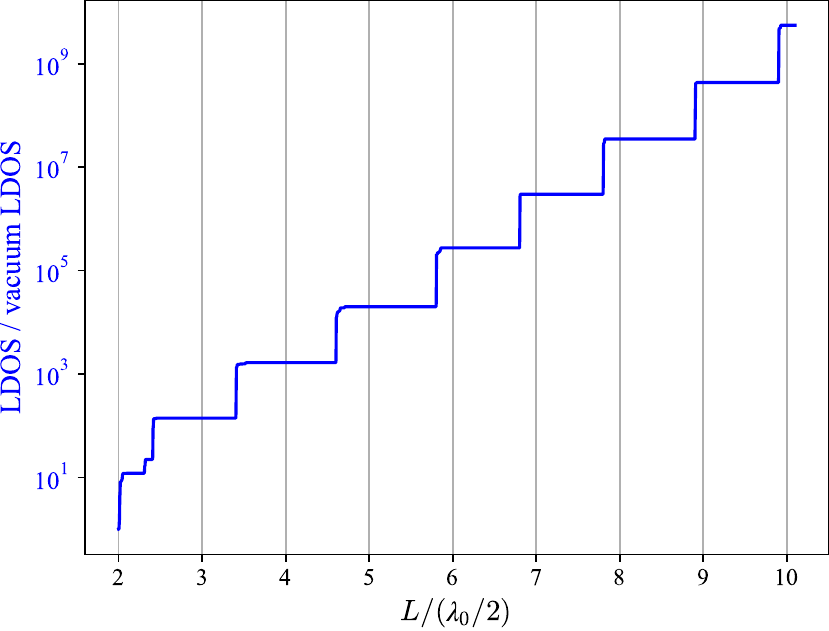}
        \label{fig:LDOS_v_L}
    \end{subfigure}
    
    % Second row: (c) 1d cavities - successive refinement
    \begin{subfigure}[t]{\textwidth}
        \centering
        \caption{}
        \includegraphics[width=\linewidth]{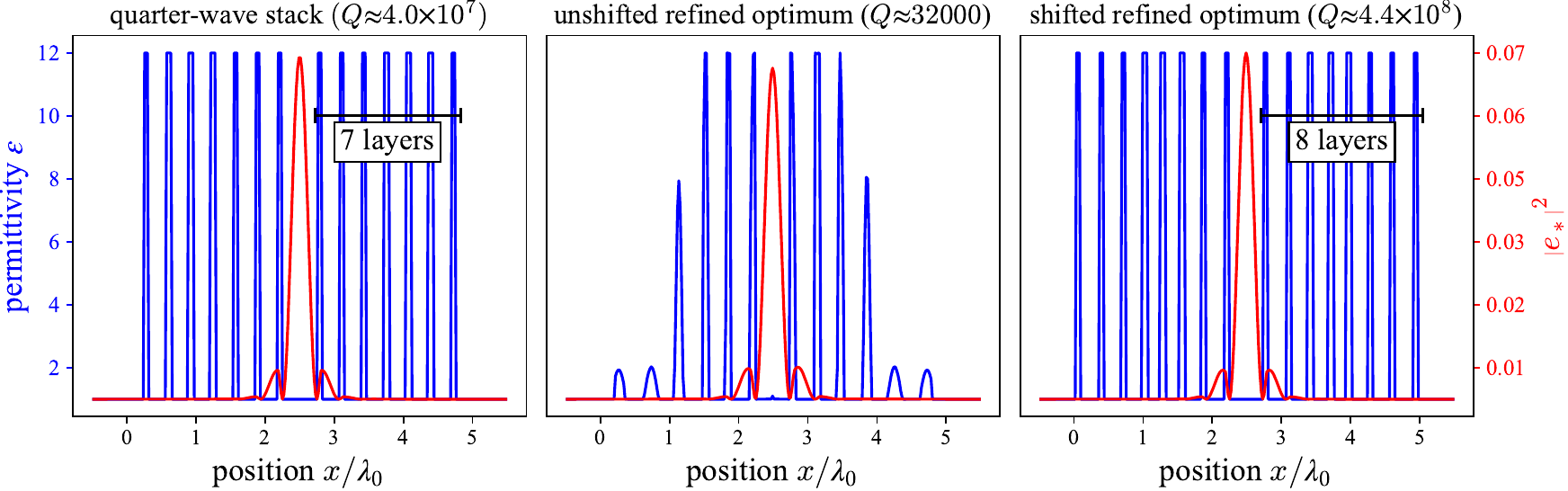}
        \label{fig:succ_ref_cavities}
    \end{subfigure}
    
    % Third row: (d) 1d plots - successive refinement
    \begin{subfigure}[t]{\textwidth}
        \centering
        \caption{}
        \includegraphics[width=\linewidth]{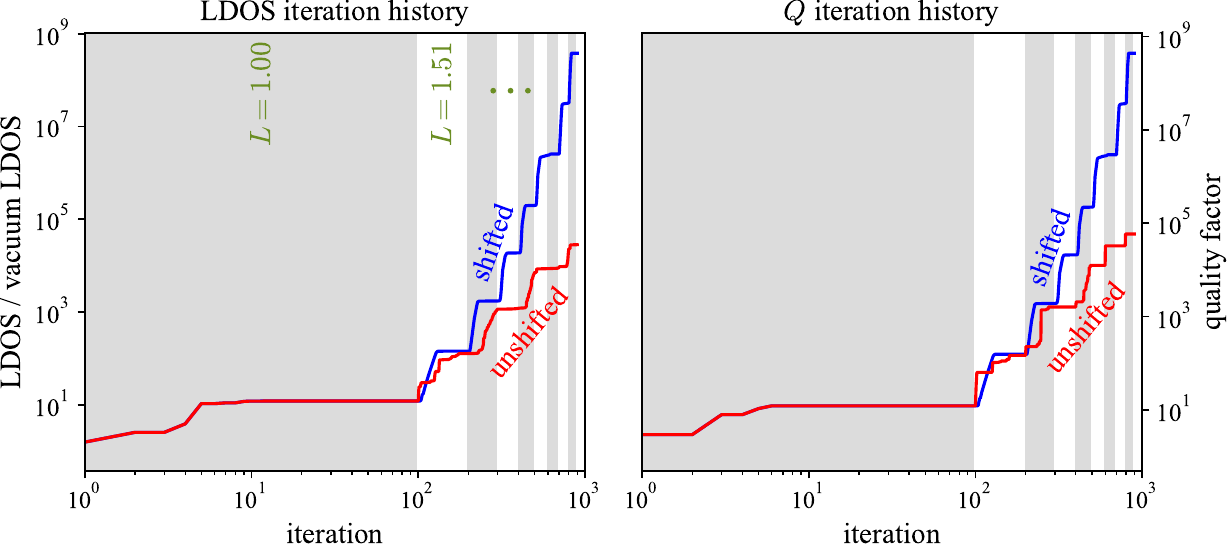}
        \label{fig:succ_ref_plots}
    \end{subfigure}
    
    \caption{(a) Schematic successive-enlargement algorithm in 1d: LDOS is optimized for a nested sequence of design domains $L_0 < L_1 < \cdots$ (green), which each optimum forming the initial guess for the next optimization. (b)~LDOS enhancement as a function of design-region size~$L$, exhibiting exponential growth in roughly half-wavelength steps. (c)~A hand-designed quarter-wave-stack cavity (left), along with optimized cavities by successive enlargement with the unshifted (middle) and shifted (right) algorithms: the resonant-mode profile (red, right axis) is superimposed over the permittivity~$\varepsilon$ (blue, left axis). (d)~LDOS enhancement (left) and $Q$ (right) as a function of the number of sparse-matrix factorizations for the successively-enlarged shifted (blue) and unshifted (red) algorithms. Alternating gray and white bands indicate a change in design-region size $L$.}
    \label{fig:succ_ref}
\end{figure}

We find that even larger $\LDOS$ and $Q$ can be obtained by a procedure of ``successive refinement'' of the degrees of freedom, a heuristic to accelerate convergence and/or evade poor local optima~\cite{MutapcicBo09}, often called ``graduated'' or ``continuation'' or ``homotopy'' optimization~\cite{Blake1987,lin2023continuation}: gradually increasing the number of design degrees of freedom, using optimization results with fewer parameters as the starting guess for optimization with more parameters.  In particular, we employ \emph{successive enlargement} of the design domain, increasing the size $L$ of the design region, using the optimum from the previous (smaller) design $L_{k}$ as the successive starting guess for $L_{k+1}$ as depicted in Fig.~\ref{fig:succ_ref_schematic}.  Our motivation for successive enlargement of the design domain in this case is evident from the structures of Fig.~\ref{fig:1d_cavities}: the optimization is clearly failing to prioritize the innermost layers of the cavity, which appear overly thin even for the shifted design, since the outer layers provide adequate confinement.

For comparison, we also considered a hand-designed quarter-wave stack~\cite{JoannopoulosJo08-book} with a $\lambda_0/2$ defect vacuum layer in the center, shown in Fig.~\ref{fig:succ_ref_cavities}~(left). It achieves $Q \approx 4.0 \times 10^7$, indicating that even the shifted design from the previous section was suboptimal.  Notice that this quarter-wave stack uses 7~high-index layers on either side of the cavity, suggesting that the previous 6-layer design is trapped in a different (topologically distinct) local optimum. 

Using successive enlargement in conjunction with the shifted algorithm, however, corrects this deficiency of inverse design.  The results are shown in Fig.~\ref{fig:succ_ref}, for which we have successively enlarged the design domain by increments of $\Delta L  = L_{k+1} - L_k  = 0.51\lambda_0$. The shifted algorithm converges to a cavity with $Q = 4.4 \times 10^8$, even better than the hand-designed cavity, in fewer than $1000$ iterations, while the resulting structure in Fig.~\ref{fig:succ_ref_cavities}~(right) is more quarter-wave-like than before (with more uniform layer thicknesses).  Moreover, it has 8 high-index layers on either side of the cavity, which is different from both the quarter-wave design and the previous non-successive optimization, suggesting the existence of multiple local optima. Successive enlargement with the unshifted algorithm yielded the same order of magnitude $Q \sim 10^4$ as the previous optimization without refinement, although the resulting unshifted cavity structure in Fig.~\ref{fig:succ_ref_cavities}~(middle) now prioritizes the innermost layers, and the optimization history suggest that it has not yet fully converged.  The convergence history, shown in Fig.~\ref{fig:succ_ref_plots}, illustrates rapid convergence of the shifted algorithm after each successive enlargement (alternating white/gray regions), whereas the unshifted algorithm is much slower.  Here, for comparison purposes we employed 100 optimization iterations per design size $L_k$; this is clearly more than was necessary for the shifted algorithm, while being insufficient for full convergence of the unshifted algorithm.  If we had halted the shifted algorithm as soon as it converged for each $L_k$, many fewer ($\ll 1000$) total iterations would be required for the same performance!

(One might wonder whether the successive optimization stages need to modify the innermost regions from the previous stages, or if it suffices to only optimize the \emph{new} portions of the design domain. We attempted a ``frozen-inner'' variant of successive enlargement in which we kept the design in the inner $L_k$  region fixed when optimizing the subsequent $L_{k+1}$ domain, only allowing $\varepsilon$ to vary in the added regions. We found that it was possible to get nearly as good LDOS and $Q$, within a factor of~$\approx 3$, but that this result was extremely sensitive to the increment $\Delta L$.)

It is also interesting to investigate the dependence of the maximum $\LDOS$ (or the $Q$) as a function of the design-region diameter $L$.  To do this, we incremented $L$ in very small steps $L_{k+1} - L_k  = 0.05 \lambda_0$, applying the shifted optimization algorithm, and plotted the resulting $\LDOS$ enhancement in Fig.~\ref{fig:LDOS_v_L}.  What we observe is that there is a minimum increment $\Delta L$ before $\LDOS$ can be substantially improved, corresponding to the minimum additional thickness before it is beneficial to begin to add another high-index layer.   The theory of quarter-wave stacks suggests that this minimum $\Delta L$ should be roughly $\lambda_0/2$, the thickness of two quarter-wave air layers (one on either side of the domain).  Indeed, we observe that the increments $\Delta L$ between successive ``steps'' in $\LDOS$ seem to be asymptotically approaching $\approx \lambda_0/2$, depicted by the gray grid lines in Fig.~\ref{fig:LDOS_v_L}.  It would be good to investigate this phenomenon more rigorously by applying recent techniques to bound the $\LDOS$ for a given design region \cite{Chao2022, Chao2025}, but this is outside the scope of the present work.  (\citeasnoun{Osting2013} also predicts an upper bound on $Q(L)$, but their bound on $\log Q$ is proportional to $L^2$ and appears to be extraordinarily loose, giving an upper bound $Q \lesssim 10^{5000}$ for $L=5\lambda_0$ with $\varepsilon \in [1,12]$.)

\subsection{Conditioning of the Hessian}
\label{sec:hessian}
The 1d optimization problem is small enough that we can explicitly compute the Hessian (second-derivative) matrix, in order to verify the predictions of Sec.~\ref{sec:second_derivative}: the unshifted $\log \LDOS$ objective should have a dominant eigenvalue scaling as  $\mathcal{O}(Q^2)$, with a corresponding eigenvector in the $\Re[\nabla_{\vec{p}}\omega_*]$ direction corresponding to frequency shifts, while the shifted objective should remove this eigenvalue. 

Computing a Hessian is well known to be costly: each column of the Hessian can be roughly as costly to compute as a single gradient computation~\cite{jax_autodiff_cookbook,matrix_calculus,griewank2008evaluating}.  In 1d with $L=5\lambda_0$, we have $n=250$ parameters ($50\,\text{pixels}/\lambda_0 \times 5\lambda_0$) , requiring $\gtrsim 250$ solves to obtain the entire Hessian even by the best known methods.  Although there are sophisticated forward-over-reverse or reverse-over-reverse methods to compute the Hessian~\cite{matrix_calculus, jax_autodiff_cookbook}, it was much easier to implement a simple finite-difference scheme that obtains the same scaling (but a worse constant factor). In particular, we applied a high-order finite-difference method, consisting of adaptive Richardson extrapolation of a centered difference~\cite[\S5.7]{Press2007Numerical}, to the gradient: each Hessian column is the partial derivative of the gradient with respect to one parameter.  Once this is computed, we remove the rows and columns of the Hessian that correspond to parameters $\vec{p}_k$ at their lower and upper bounds $\{0,1\}$, since variables with active bound constraints are effectively removed from the optimization problem. The remaining $\approx 200 \times 200$ Hessian is then diagonalized and we examine its eigenvalues and eigenvectors.

\begin{figure}[tbp]
    \centering
    \includegraphics[width=0.8\linewidth]{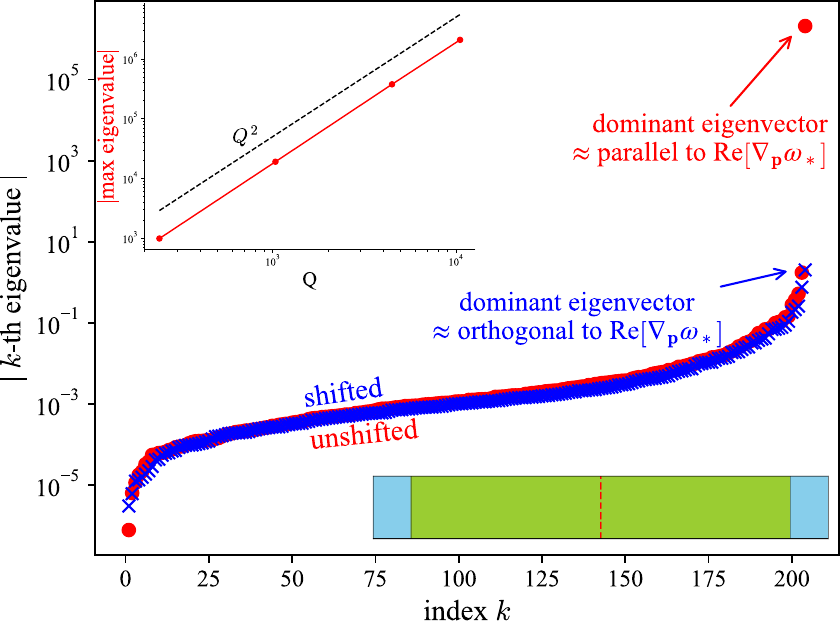}
    \caption{Eigenvalues of the Hessians for the shifted (blue crosses) and unshifted (red dots) objective functions, computed at the unshifted initialization of our 1d test case (lower-right inset) from Fig.~\ref{fig:1d_cavities}(left). The largest eigenvalues differ by six orders of magnitude. Upper-left inset verifies that the dominant eigenvalue of the unshifted objective (red dots) scales like $Q^2$ (dashed black line), as predicted in Sec.~\ref{sec:second_derivative}.}
    \label{fig:spectrum}
\end{figure}

The result, applied to the unshifted initialization of Fig.~\ref{fig:1d_cavities}~(left) with $Q \approx 11000$, is shown in Fig.~\ref{fig:spectrum}. As expected, the Hessian of the unshifted $\log \LDOS$ objective, whose eigenvalues are shown as red circles, is dominated by a single eigenvalue that is about $10^5 \times$ the next-biggest eigenvalue.  We find that the corresponding eigenvector is almost exactly parallel to $\Re[\nabla_{\vec{p}}\omega_*]$ (differing by $< 10^{-4}$~degrees).  To obtain the scaling with $Q$, shown in inset of Fig.~\ref{fig:spectrum}, we applied the same procedure at 50, 100, 150, and 200 iterations of the unshifted algorithm starting from vacuum (yielding $Q\approx 240, 1000,4500,11000$, respectively); the resulting maximum Hessian eigenvalue scales almost exactly proportional to $Q^2$, as predicted.  In contrast, when we compute the eigenvalues of the \emph{shifted} $\log \LDOS$ objective, shown as blue $\times$'s in Fig.~\ref{fig:spectrum}, the largest eigenvalue of the unshifted Hessian is removed.  The remaining dominant eigenvalue ($\sim 10^5$ times smaller) corresponds to an eigenvector that is nearly \emph{perpendicular} to $\Re[\nabla_{\vec{p}}\omega_*]$  (angle $\approx 86^\circ$).

\section{Two-dimensional results}
\label{sec:2d_results}
In this section, we apply the unshifted and shifted optimization algorithms to a two-dimensional cavity, considering the challenging case of coupling to a point source that lies just \emph{outside} the design region (so that a cavity cannot simply surround the source with a photonic-bandgap material~\cite{Liang2013,JoannopoulosJo08-book}).  This scenario was also considered in recent theoretical upper bounds~\cite{Chao2022,Chao2025}.

\begin{figure}[tbp]
    \centering
    \includegraphics[width=0.54\linewidth]{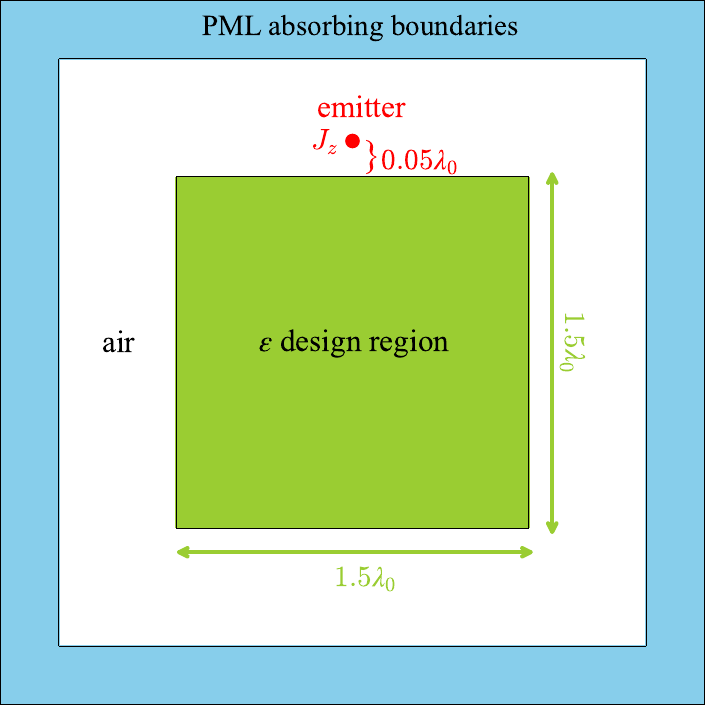}
    \caption{Schematic example LDOS-optimization problem in 2d: a square $1.5\lambda_0 \times 1.5\lambda_0$ design region (green), surrounded by air ($\varepsilon=1$) and PML absorbers (blue) is optimized to maximize the LDOS of a point source (red dot) $0.05\lambda_0$ outside the design region.}
    \label{fig:2d_region}
\end{figure}

In particular, Fig.~\ref{fig:2d_region}. depicts our two-dimensional design region $\Omega$, a $1.5\lambda_0 \times 1.5\lambda_0$ square region padded on all sides by $0.5\lambda_0$ of air and surrounded by $0.25\lambda_0$ of PML. The (out-of-plane) source is centered at a distance $0.05\lambda_0$ above the upper edge of $\Omega$. The discretization has a resolution of $20\,\text{pixels}/\lambda_0$.

\begin{figure}[tbp]
    \centering
    
    % Top row: (a) 2d cavities
    \begin{subfigure}[t]{\textwidth}
        \centering
        \caption{}
        \includegraphics[width=\linewidth]{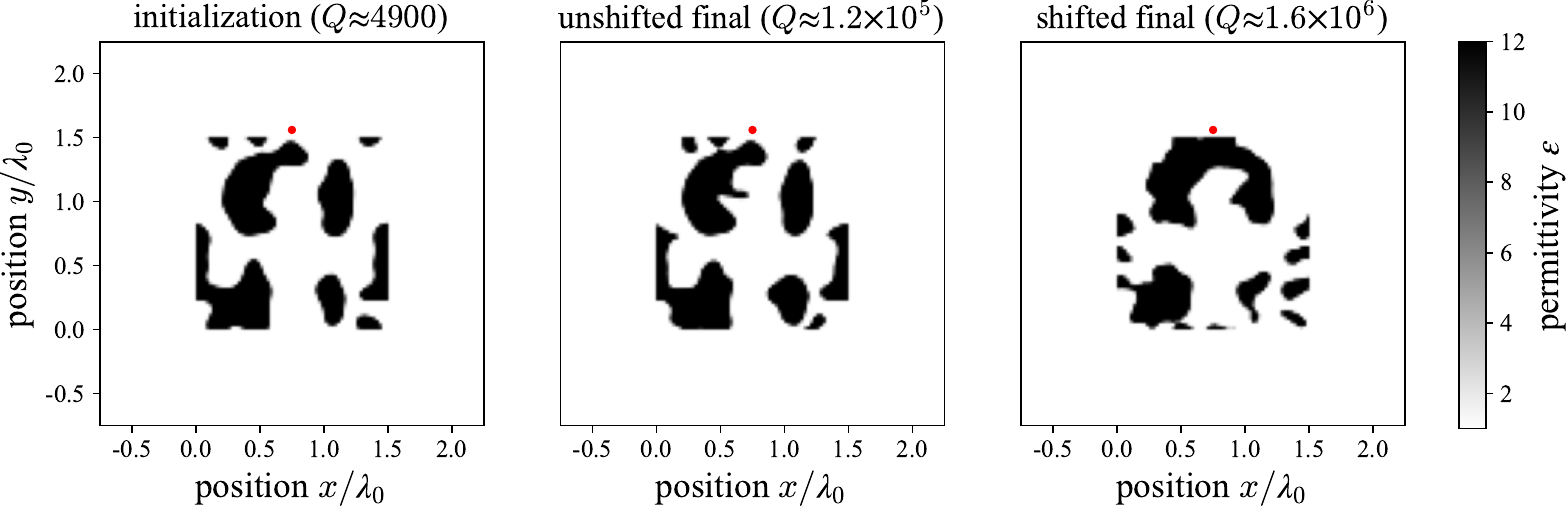}
        \label{fig:2d_cavities}
    \end{subfigure}
    
    % Second row: (b) 2d modes
    \begin{subfigure}[t]{\textwidth}
        \centering
        \caption{}
        \includegraphics[width=\linewidth]{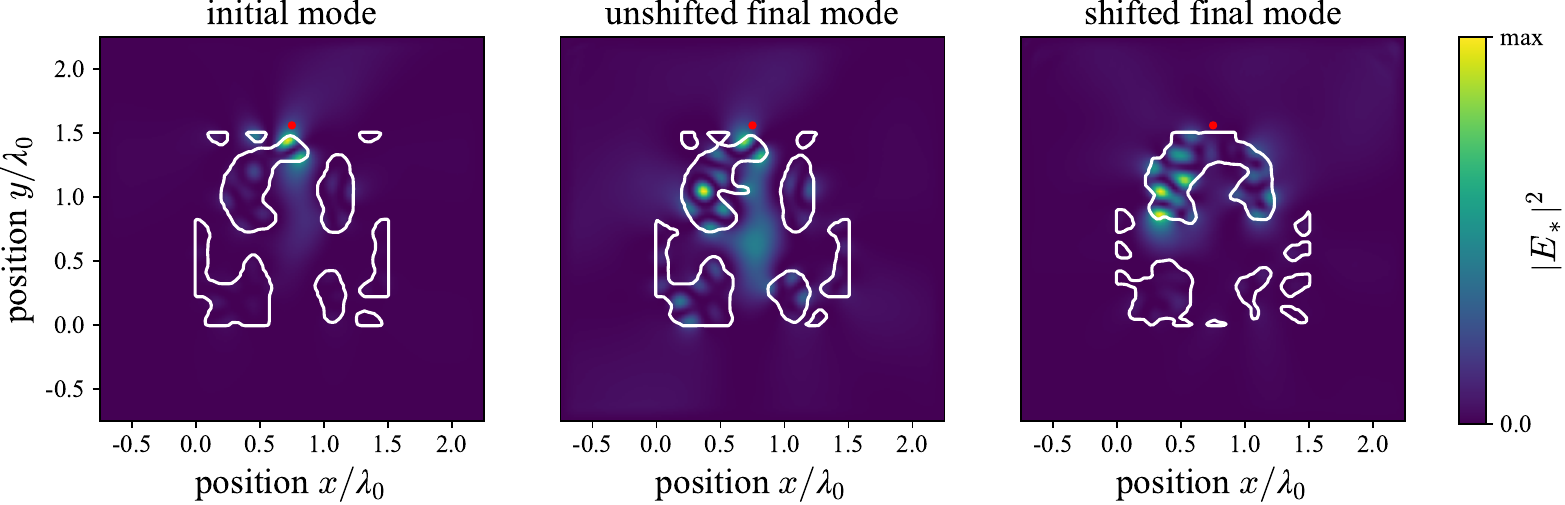}
        \label{fig:2d_modes}
    \end{subfigure}
    
    % Third row: (c) 2d plots
    \begin{subfigure}[t]{\textwidth}
        \centering
        \caption{}
        \includegraphics[width=\linewidth]{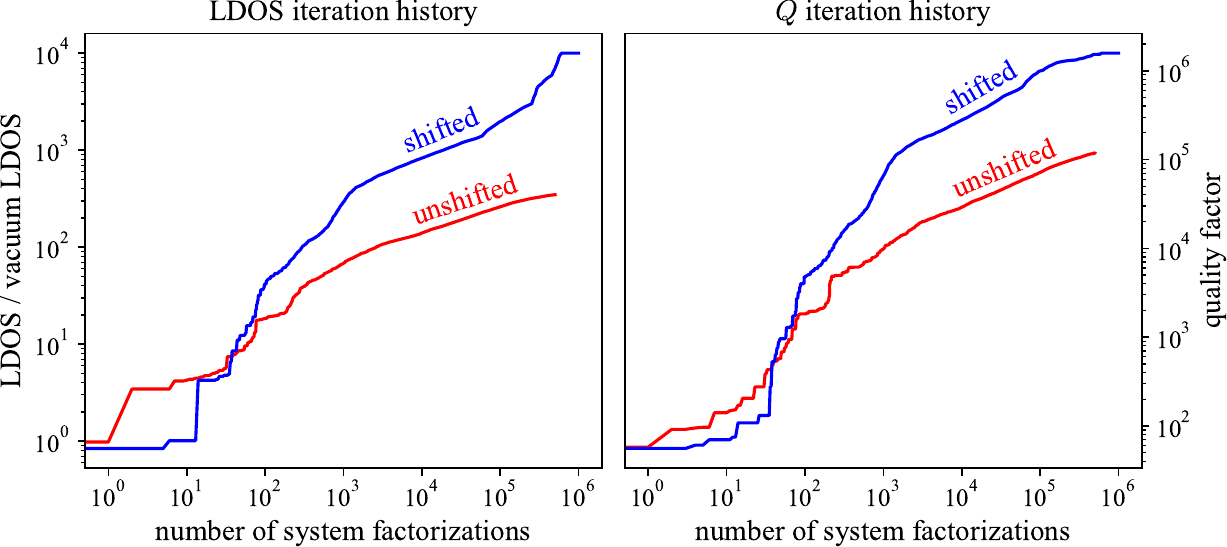}
        \label{fig:2d_plots}
    \end{subfigure}
    
    \caption{(a) Unshifted initialization after 1000 iterations (left) and shifted (right) and unshifted (middle) final cavities after 500,000 iterations, with the source location (from Fig.~\ref{fig:2d_region}) shown as a red dot. (b) Resonant-mode electric-field intensities  for the initialization (left) and unshifted (middle) and shifted (right) final cavities, with the interface between air ($\varepsilon = 1$) and dielectric ($\varepsilon = 12$) superimposed in white. (c) LDOS enhancement (left) and $Q$ (right) as a function of the number of sparse-matrix factorizations for the shifted (blue) and unshifted (red) algorithms.}
\end{figure}

Starting from a uniform intermediate permittivity $\varepsilon = 6.5$, we obtained the unshifted initialization shown in Fig.~\ref{fig:2d_cavities}~(left), by running 1000 iterations of the unshifted algorithm. As explained in Sec.~\ref{sec:topopt}, we applied conic filter radius of $0.1\lambda_0$ to set a minimum lengthscale in the design, and binarized with subpixel-smoothed projection at a sequence of increasing binarization strengths $\beta$: 100 iterations at $\beta=8$, 200 at $\beta=16$, 300 at $\beta=40$, and 400 at $\beta=\infty$. From this initialization, both the shifted and unshifted algorithms were run for 500,000 iterations. The shifted algorithm converged to a design with $Q \approx 1.6\times 10^6$, while the unshifted algorithm was still not fully converged at $Q \approx 1.2\times 10^5$, as shown in Fig.~\ref{fig:2d_cavities}~(middle, right). The convergence histories for both LDOS and $Q$ are plotted in Fig.~\ref{fig:2d_plots}, demonstrating that the shifted algorithm again exhibits substantially faster convergence and higher final performance.  (As in Sec.~\ref{sec:1d_results}, we plot convergence versus the number of Maxwell-matrix factorizations in order to account for the fact that the shifted iterations are roughly twice as expensive.) For example, the shifted algorithm reaches a $Q$  of $10^5$ in $\approx 1000\times$ fewer iterations than the unshifted algorithm. The result resonant-mode $| E_*|^2$ fields are shown Fig.~\ref{fig:2d_modes}: they exhibit a tradeoff between strong spatial localization near the source and strong temporal confinement (high~$Q$).

% TODO: compare to ring resonator? TODO: successive enlargement?

\subsection{Comparison to upper bounds}
It is also interesting to compare to the theoretical upper bounds derived for a similiar geometry~\cite{Chao2025}. These bounds were applied to a slightly different material $\varepsilon=6+10^{-4}i$ that includes absorption loss ($\omega \Im \varepsilon > 0$), which helps limit the attainable $Q$. In \citeasnoun{Chao2025}, inverse design struggled to approach the theoretical upper bound, potentially due to a combination of slow convergence for high $Q$ and being trapped in a poor local minimum: for an $L \times L$ design region with $L=3\lambda_0$, separated by the source at a distance $0.2\lambda_0$, unshifted inverse design  from various random and deterministic initializations (including vacuum) with 15,000 iterations yielded an an $\LDOS$ that was $\approx 5\times$ below the theoretical bound.  However, \citeasnoun{Chao2025} showed that the upper-bound procedure could be exploited to predict a good initial guess for inverse design, from which unshifted inverse design yielded an $\LDOS$   $\approx 1.6\times$ below the theoretical bound.  Now that our shifted algorithm is not limited by slow convergence, we can apply it to the same problem and characterize the influence of the starting guess with greater certainty.

Applying our shifted optimization to the same $3\lambda_0 \times 3\lambda_0$, $0.2\lambda_0$~separation, $\varepsilon=6+10^{-4}i$   design problem, with a simple \emph{vacuum} initial structure followed by the same 1000-step unshifted initialization as above, we found that it converged within 20,000 shifted iterations. The resulting $\LDOS$ is about $2.9\times $ below the theoretical bound.  The corresponding lifetime is $Q \approx 39000$.  This is significantly better than the previous unshifted inverse designs from similar starting points, suggesting that those previous results were limited by the slow convergence, although the $Q$  in this case is small enough to practically reach with a larger number of iterations.  However, it is still worse than optimization from a more sophisticated starting structure, suggesting a suboptimal local minimum. 

\section{Conclusion}
\label{sec:conclusion}
Our results show that the new shifted LDOS-optimization algorithm can converge \emph{orders of magnitude} faster than previous unshifted algorithms for $Q \gg 1000$ structures. Like any nonconvex optimization, it can still obtain a suboptimal local minimum, but this problem can be ameliorated by improved starting structures obtained from theoretical bounds~\cite{Chao2025} or by heuristics such as successive enlargement of the design region. The drastic reduction in the number of optimization steps makes the shifted algorithm beneficial even though each evaluation of the objective function is more expensive (requiring an eigensolve), especially if the dominant cost is an explicit sparse-matrix factorization as in most previous resonance-optimization works: the shifted algorithm only doubles the number of factorizations compared to the unshifted method.

The same shifted algorithm should be applicable to other resonant-response objectives that have been optimized in previous work, such as nonlinear responses involving integrals of $\Vert \vec{E}\Vert^4$~\cite{Yao2023,Martinez2025} or compositions of multiple linear responses~\cite{Mann2023, LinLi16, Christiansen2020, Pan2021}. One should simply compute the \emph{same} objective function at a different frequency given by the nearest resonant mode, after initializing with a few iterations of the unshifted algorithm to identify the resonance. There are also several other ways in which our shifted algorithm could potentially be improved. Rather than re-computing the $\LDOS$  or other objective at the shifted frequency, one could attempt to approximate the objective directly in terms of the computed resonant mode~\cite{OskooiJo13-sources}, but this would make gradient computation more complicated and costly: gradients of eigenvector-dependent objectives typically require the solution of an additional linear system~\cite{He2023,Li2024}. Another opportunity is to find better algorithms to prevent the eigensolver from ``jumping'' to a different nearby resonance, perhaps by monitoring changes in the eigenvector; currently, we have found that it suffices to impose a simple lower or upper bound on $\Re \omega_*$, but some experimentation may currently be required for each new system. Finally, it would also be valuable to devise algorithms that address \emph{other} sources of slow convergence besides frequency shifts. For example, we found in 1d that successive enlargement greatly accelerated convergence in \emph{addition} to escaping a poor local optimum, and it would interesting to determine why this acceleration occurs and whether related strategies can yield similar gains elsewhere.

\section*{Appendix A: Gradient computation}

In this section, we derive the gradients of the unshifted and shifted objective functions with respect to $\vec{\varepsilon}$.  (These derivatives can then be propagated to the underlying ``density'' degrees of freedom via a straightforward chain rule, for which we applied automatic differentiation~\cite{JMLR:v18:17-468,Zygote.jl-2018}.)  The key goal is to obtain an ``adjoint'' (or ``reverse-mode'') formulation in which computing the gradient has roughly the same cost as evaluating the objective function \emph{once}, regardless of the number of parameters~\cite{JMLR:v18:17-468,griewank2008evaluating,matrix_calculus,Hughes2018,Molesky2018}. In fact, we find an even better result in both the unshifted and shifted cases: the gradient computation incurs \emph{negligible additional cost} compared to the computation of the LDOS itself. To reduce the chance of an algebra error, all of our gradient formulas were validated numerically against a finite-difference approximation.

\subsection{The unshifted gradient}
\label{sec:unshifted_grad}
First, let us review the gradient calculation for unshifted $\LDOS$~\cite{Liang2013}. For all derivations that follow, let $\vec{u}^{(k)}$ denote the $k$-th Cartesian basis vector; this is useful because our discretized Maxwell matrix $\mat{A}$ only depends on $\varepsilon_k$ (the permittivity at pixel $k$) in the $k$-th diagonal element, so $\frac{\partial \mat{A}}{\partial \vec{\varepsilon}_k}=-\omega_0^2 \vec{u}^{(k)} (\vec{u}^{(k)})^\T$ (a rank-1 matrix with only a single nonzero entry). We will also use the reciprocity identity~\cite{Chew2008,Guo2022} $\mat{A}^T=\mat{A}$ (enforced for our discretization by Appendix~B). If $\vec{e} = \mat{A}^{-1} \vec{b}$, for a real vector $\vec{b}$ ($=\vec{u}^{(j)}$ at the source location $j$), the chain rule then yields~\cite{Liang2013,matrix_calculus}:
\begin{align*}
    \frac{\partial \vec{b}^{\dag} \vec{e}}{\partial \vec{\varepsilon}_k} &= \vec{b}^{\T}\frac{\partial \mat{A}^{-1}}{\partial \vec{\varepsilon}_k}\vec{b}\\
    &= -\vec{b}^{\T}\mat{A}^{-1}\frac{\partial \mat{A}}{\partial \vec{\varepsilon}_k}\mat{A}^{-1}\vec{b}\\
    &= -\vec{b}^{\T}\mat{A}^{-1}\left(-\omega_0^2 \vec{u}^{(k)}\left(\vec{u}^{(k)}\right)^{\T}\right)\mat{A}^{-1}\vec{b}\\
    &= \omega_0^2 \vec{e}^{\T}\vec{u}^{(k)}\left(\vec{u}^{(k)}\right)^{\T}\vec{e}\\
    &= \omega_0^2 \vec{e}_k^2 \, ,
\end{align*}
where in the second-to-last line we employed reciprocity.  Thus, since $\LDOS = -\Im[\vec{e}^\dag \vec{b}] = \Im[\vec{b}^\dag \vec{e}]$, we have that
\begin{equation}
\label{eq:unshifted_grad_computation}
    \frac{\partial \LDOS}{\partial \vec{\varepsilon}_k} = \omega_0^2 \Im\left[\vec{e}_k^2\right] \, ,
\end{equation}
which can be computed for all components $k$ with \emph{no} additional Maxwell solves beyond the solve for $\vec{e}$ required to compute the $\LDOS$ itself: the gradient is essentially ``free.''

\subsection{The shifted gradient}
\label{sec:shifted_grad}
We are now letting the frequency be a function of $\vec{p}$ in our LDOS calculation:
\begin{equation}
\label{eq:shifted_formulation_appendix}
    \LDOS := \LDOS(\Re\left[\omega_*(\vec{p}, \omega_0)\right], \vec{x}_0, \vec{p})\, .
\end{equation}
This leads to an additional term in the chain rule, since the $\LDOS$ now depends on $\vec{p}$ through both $\varepsilon$ and $\omega$:
\begin{equation}
\label{eq:shifted_grad_theory}
    \frac{\mathrm{d}\LDOS}{\mathrm{d}\vec{p}_k} = \left. \frac{\partial\LDOS}{\partial \vec{p}_k}\right|_{\omega = \Re\omega_*} + \left(\left.\frac{\partial\LDOS}{\partial \omega}\right|_{\omega = \Re\omega_*}\right) \Re\left[\frac{\partial \omega_*}{\partial \vec{p}_k}\right]\, .
\end{equation}
The first term is given by Eq.~\ref{eq:unshifted_grad_computation}. The partial derivative with respect to $\omega$ can be computed as follows:
\begin{align*}
    \frac{\partial \vec{b}^{\dag} \vec{e}}{\partial \omega} &= \vec{b}^{\T}\frac{\partial \mat{A}^{-1}}{\partial \omega}\vec{b}\\
    &= -\vec{b}^{\T}\mat{A}^{-1}\frac{\partial \mat{A}}{\partial \omega}\mat{A}^{-1}\vec{b}\\
    &= -\vec{b}^{\T}\mat{A}^{-1}(-2\omega \mat{D})\mat{A}^{-1}\vec{b}\\
    &= 2\omega
    \vec{e}^{\T}\mat{D}\vec{e} \, .
\end{align*}
Thus,
\begin{equation}
\label{eq:partial_omega}
    \left.\frac{\partial\LDOS}{\partial \omega}\right|_{\Re\omega_*} = 2\Re\omega_*\cdot\Im\left[\vec{e}^{\T}\mat{D}\vec{e}\right]\, ,
\end{equation}
which (as for $\partial \LDOS / \partial \varepsilon_k$ above) is essentially free: we re-use $\vec{e}$  (from the Maxwell solve at the frequency $\Re \omega_*$), and $\mat{D}$ is diagonal.

Now, let $\mat{M} = \mat{D}^{-1}\mat{A}_0$, where $\mat{A}_0 = \mat{A}(\omega_0)$, and let $\vec{e}_*$ and $\vec{f}_*$ be the right and left eigenvectors, respectively, of $\mat{M}$ with eigenvalue $\mu_* = \omega_*^2 - \omega_0^2$. That is, $\mat{M}\vec{e}_* = \mu_*\vec{e}_*$ and $\mat{M}^{\dag}\vec{f}_* = \overline{\mu_*}\vec{f}_*$. From eigenvalue perturbation theory~\cite{Horn2012}, we know that
\begin{equation}
\label{eq:perturbation_theory}
    \frac{\partial \mu_*}{\partial \vec{\varepsilon}_k} = \frac{\partial \omega_*^2}{\partial \vec{\varepsilon}_k} = \frac{\left(\vec{f}_*\right)^{\dag}\frac{\partial \mat{M}}{\partial \vec{\varepsilon}_k}\vec{e}_*}{\left(\vec{f}_*\right)^{\dag}\vec{e}_*}\, .
\end{equation}
(We computed $\vec{e}_*$ using the shift-and-invert Arnoldi iteration~\cite{bai2000templates,montoison-orban-2023} applied to $\mat{M}^{-1}$, which requires only \emph{one} additional sparse-matrix factorization of our Maxwell matrix $\mat{A}_0$ at $\omega_0$.) By the symmetry of $\mat{A}$ and $\mat{D},$ it follows that $\overline{\vec{f}_*} = \mat{D}\vec{e}_*$. In particular, combining this vector with the identity  $\mat{M}^{\T} = \mat{D}\mat{M}\mat{D}^{-1}$ verifies the left eigen-equation:
\begin{equation}
\label{eq:check_left_eigvec}
    \mat{M}^{\T}\overline{\vec{f}_*} = \mat{D}\mat{M}\mat{D}^{-1}\overline{\vec{f}_*} = \mat{D}\mat{M}\vec{e}_* = \mu_*\mat{D}\vec{e}_* = \mu_*\overline{\vec{f}_*}\, .
\end{equation}
We can now substitute into Eq.~\ref{eq:perturbation_theory} to obtain
    $$\frac{\partial \omega_*^2}{\partial \vec{\varepsilon}_k} = \frac{\vec{e}_*^{\T}\mat{D}\frac{\partial \mat{M}}{\partial \vec{\varepsilon}_k}\vec{e}_*}{\vec{e}_*^{\T}\mat{D}\vec{e}_*},$$
where
\begin{align*}
    \frac{\partial \mat{M}}{\partial \vec{\varepsilon}_k} &= \frac{\partial}{\partial \vec{\varepsilon}_k}\left(\mat{D}^{-1}\mat{A}\right)\\
    &= \frac{\partial}{\partial \vec{\varepsilon}_k}\left(-\mat{D}^{-1}\mat{L} - \omega_0^2\mat{I}\right)\\
    &= -\frac{\partial}{\partial \vec{\varepsilon}_k}\left(\mat{D}^{-1}\right)\mat{L}\\
    &= \vec{\varepsilon}_k^{-2}\vec{u}^{(k)}\left(\vec{u}^{(k)}\right)^{\dag}\mat{L} \, .
\end{align*}
Here, $\mat{L}$ is the discretized Laplacian (from Sec.~\ref{sec:LDOS}).  Therefore, we have
\begin{align*}
    \frac{\partial \omega_*^2}{\partial \vec{\varepsilon}_k} &= \vec{\varepsilon}_k^{-2}\frac{\vec{e}_*^{\T}\mat{D}\vec{u}^{(k)}\left(\vec{u}^{(k)}\right)^{\dag}\mat{L}\vec{e}_*}{\vec{e}_*^{\T}\mat{D}\vec{e}_*}\\
    &= -\vec{\varepsilon}_k^{-2}\omega_*^2\frac{\vec{e}_*^{\T}\mat{D}\vec{u}^{(k)}\left(\vec{u}^{(k)}\right)^{\dag}\mat{D}\vec{e}_*}{\vec{e}_*^{\T}\mat{D}\vec{e}_*}\\
    &= -\omega_*^2\frac{\left(\vec{e}_*\right)_k^2}{\vec{e}_*^{\T}\mat{D}\vec{e}_*} \, .
\end{align*}
We therefore obtain the last term required for the computation of the $\LDOS$ derivative above:
\begin{equation}
\label{eq:gradient_of_omega}
    \frac{\partial \omega_*}{\partial \vec{\varepsilon}_k} = -\frac{\omega_*}{2} \frac{\left(\vec{e}_*\right)_k^2}{\vec{e}_*^{\T}\mat{D}\vec{e}_*} \, .
\end{equation}
The computational cost of the shifted LDOS is roughly twice that of the unshifted LDOS, assuming that direct sparse factorization~\cite{davis2006direct} of $\mat{A}$ is employed, because the cost of this factorization typically dominates the computation time. That is, one now requires two factorizations: one at frequency $\omega_0$ to use with the shift-invert eigensolver, and one at the new frequency $\Re \omega_*$.   However, given the shifted-LDOS computation, the gradient of the shifted LDOS is again essentially free, since it involves only simple computations with $\vec{e}$ and $\vec{e}_*$, with a total computation time that scales linearly with the number of degrees of freedom (the number of pixels), much faster than sparse-matrix factorizations or solves.

\section*{Appendix B: Preserving reciprocity $\mat A=\mat A{}^{\T}$ with PML}

Electromagnetic reciprocity implies that the linear operator relating
fields and currents (or its inverse, the dyadic Green's function)
is \emph{complex-symmetric} assuming reciprocal materials $\varepsilon=\varepsilon^{\T}$
and $\mu=\mu^{\T}$ (but is not generally Hermitian in the presence
of loss or outgoing boundary conditions)~\cite{Chew2008,Guo2022}. For the algorithms in this
paper, it is convenient to ensure that the corresponding matrix property $\mat A=\mat A{}^{\T}$
is preserved when the equations are discretized---this yields
a simple identity between the left and right eigenvectors, both of
which are needed for gradient computation (see Appendix~A), so that
they need not be computed separately. However, this symmetry depends
on how one formulates the perfectly matched layer (PML)~\cite{taflove_em_book,pml_notes} that implements
the absorbing/outgoing boundaries. In this Appendix, we discuss how
one can formulate the PML to preserve reciprocity.

PML can be elegantly derived from the ``stretched-coordinate'' viewpoint,
in which one first analytically continues the equations to \emph{complex}
coordinates (in which waves become exponentially decaying without
creating reflections), and then one performs a change of variables
back to real coordinates~\cite{Chew1994,taflove_em_book,pml_notes}. For a PML along the $x$ direction,
this transforms every $\frac{\partial}{\partial x}$ into $\frac{1}{s_{x}(x)}\frac{\partial}{\partial x}$
, with a complex ``stretch'' factor is typically~\cite{taflove_em_book,pml_notes,OskooiJo11} $s_{x}(x)=1+\frac{i\sigma_{x}(x)}{\omega}$,
where $\sigma_{x}=0$ in the interior of the domain (where solutions
are unchanged) and $\sigma_{x}/\omega>0$ in the PML absorbing layers
(inducing reflectionless attenuation) where $\sigma_x$ ramps up gradually (often quadratically) to counteract discretization effects~\cite{taflove_em_book,pml_notes,OskooiJo11}. Similarly, $\frac{\partial}{\partial y}$
is replaced by $\frac{1}{s_{y}(x)}\frac{\partial}{\partial y}$. In
our 2d Helmholtz equation~\eqref{eq:Helmholtz_continuous}, this corresponds to the transformation 
\begin{equation}
\nabla^{2}+\omega^{2}\varepsilon\longrightarrow\frac{1}{s_{x}}\frac{\partial}{\partial x}\frac{1}{s_{x}}\frac{\partial}{\partial x}+\frac{1}{s_{y}}\frac{\partial}{\partial y}\frac{1}{s_{y}}\frac{\partial}{\partial y}+\omega^{2}\varepsilon\,.\label{eq:PML-stretched}
\end{equation}
Unfortunately, this formulation breaks reciprocity (and leads to an
asymmetric matrix $\mat A$), because it has scale factors
$s_{x}^{-1}$ and $s_{y}^{-1}$ on the left but not on the right. One can restore the symmetry with a diagonal similarity transformation
$\mat A\to\mat S\mat A\mat S^{-1}$ (a change of basis), however,
because $s_{y}$ commutes with $\frac{\partial}{\partial x}$ and
$s_{x}$ commutes with $\frac{\partial}{\partial y}$. In particular,
one can multiply on the left by $\sqrt{s_{x}s_{y}}$ and divide on
the right by $\sqrt{s_{x}s_{y}}$, yielding the symmetric operator
\begin{equation}
\frac{1}{\sqrt{s_{x}}}\frac{\partial}{\partial x}\frac{1}{s_{x}}\frac{\partial}{\partial x}\frac{1}{\sqrt{s_{x}}}+\frac{1}{\sqrt{s_{y}}}\frac{\partial}{\partial y}\frac{1}{s_{y}}\frac{\partial}{\partial y}\frac{1}{\sqrt{s_{y}}}+\omega^{2}\varepsilon\,.\label{eq:PML-symmetrized}
\end{equation}
This is the PML formulation used for the results in this paper, which we validated by the method of \citeasnoun{OskooiJo11}. Note
that such transformations do not change the eigenvalues and
merely scale the eigenvectors by $\sqrt{s_{x}s_{y}}$ (as well as
scaling the right-hand side currents by $1/\sqrt{s_{x}s_{y}}$), which
only changes the resonant mode profile and currents inside the PML
(since $\sqrt{s_{x}s_{y}}=1$ in the interior of the domain). Hence,
it has no effect on figures of merit like $\LDOS$ that depend only
on the interior fields (and use currents supported only in the interior),
nor does it affect our sensitivity analysis of Appendix~A. This transformation only changes the Laplacian matrix $\mat L$ and
not the diagonal material matrix $\mat D$ in Eq.~\eqref{eq:Helmholtz_discrete}.

An alternative transformation that would accomplish the same goal
of preserving reciprocity is to simply multiply both sides the field--current
relation $\mat A\vec{e}=\vec{b}$ or the eigen-equation $\mat A\vec{e}=\vec{0}$
on the left by the diagonal operator (hence diagonal matrix) $s_{x}s_{y}$,
yielding the operator
\begin{equation}
\frac{\partial}{\partial x}\frac{s_{y}}{s_{x}}\frac{\partial}{\partial x}+\frac{\partial}{\partial y}\frac{s_{x}}{s_{y}}\frac{\partial}{\partial y}+\omega^{2}s_{x}s_{y}\varepsilon\,,\label{eq:UPML}
\end{equation}
which is also complex-symmetric, and again transforms the currents only in the PML regions (since $s_{x}s_{y}=1$ in the interior),
but affects both $\mat L$ and $\mat D$. (This is not technically
a similarity transformation because we only multiplied on the left,
but it does not change the matrix-pencil/generalized-eigenproblem solutions
of $\mat A\vec{e}=\vec{0}$.) It is straightforward to derive that
this corresponds exactly to the ``UPML'' formulation of PML~\cite{taflove_em_book,OskooiJo11},
in which the Jacobian scale factors of the stretched-coordinate formulation
are converted into a modification of the fields and materials in the
ordinary Maxwell equations by transformation optics~\cite{OskooiJo11}. In
particular, the PML ``stretch'' factors correspond (in 3d) to a $3\times3$
Jacobian matrix 
\begin{equation}
\mathcal{J}=\begin{pmatrix}s_{x}^{-1}\\
 & s_{y}^{-1}\\
 &  & s_{z}^{-1}
\end{pmatrix}\label{eq:PML-jacobian}
\end{equation}
and the materials $\varepsilon$ and $\mu$ are transformed by~\cite{OskooiJo11}
\begin{equation}
\varepsilon\to\frac{\mathcal{J}\varepsilon\mathcal{J}^{\T}}{\det\mathcal{J}}\,,\;\mu\to\frac{\mathcal{J}\mu\mathcal{J}^{\T}}{\det\mathcal{J}}\,,\label{eq:transformation-optics}
\end{equation}
which obviously preserves reciprocity $\varepsilon=\varepsilon^{\T}$
and $\mu=\mu^{\T}$. Applied to the 3d Maxwell operator $-\nabla\times\mu^{-1}\nabla\times+\omega^{2}\varepsilon$
and simplified to the 2d case
for $z$-invariant $E_{z}$-polarized fields with $\mu=1$ and scalar $\varepsilon$, this yields
Eq.~(\ref{eq:UPML}) above. (For iterative solvers, it has been shown
that the stretched-coordinate formulation is more efficient than the
UPML formulation~\cite{Shin2012}, but one can always
transform between formulations after the solver is complete. For sparse-direct solvers, all of these formulations, which differ only by diagonal
scale factors, have similar costs.)

\begin{backmatter}

\bmsection{Funding}
This work was supported in part by the U.S. Army Research Office through the Institute for Soldier Nanotechnologies (award no.~W911NF-23-2-0121), by the Simons Foundation through the Simons Collaboration on Extreme Wave Phenomena, by the MIT Undergraduate Research Opportunities Program, and by the Novo Nordisk Foundation Quantum Computing Programme  (award no.~NNF22SA0081175).

\bmsection{Disclosures}
The authors declare no conflicts of interest.

\bmsection{Data availability}
Data underlying the results presented in this paper are not publicly available at this time but may be obtained from the authors upon reasonable request.

\end{backmatter}

\bibliography{urop_refs}

\end{document}